\documentclass{JHEP3}

\usepackage{epsfig}
\usepackage{feynmp}

\title{The Relic Abundance of Massive Colored Particles after a Late Hadronic Annihilation Stage}
\author{Chen Jacoby and Shmuel Nussinov\\
Raymond and Beverly Sackler Faculty of Exact Sciences \\
School of Physics and Astronomy \\
Tel-Aviv University, Ramat-Aviv 69978, Israel \\
E-mail: \email{chj3@post.tau.ac.il},
\email{nussinov@post.tau.ac.il}}

\abstract{We discuss the relic abundance of massive long lived colored particles with mass of the order of $1$ TeV. We first examine the case where the massive colored particles have the standard
color only. Next we consider the ``Quirk Model'' suggested by M. Luty, in which the colored particles transform under an additional non-abelian gauge group with a scale much smaller than the particles' mass. In both cases, the relic abundance is reduced via a ``late'' hadronic annihilation stage. In the second case the relic Quirks bind to ordinary quarks forming fractional charged objects and also anomalous heavy isotopes, and the bounds on the relic abundance become extremely severe. The force between Quirks, however, has a new confining part that manifests via macroscopic strings and the resulting efficient ``very late'' annihilations reduce the relic abundance to acceptable levels. The prospects of creating and detecting such particles at LHC and the fate of the particles created are discussed.}


\preprint{arXiv:0712.2681\\
TAUP-2867-07}

\begin{document}
\section{The Reduced Relic Abundance of Massive Long Lived Colored Particle} \label{sec:reduced}

New physics of some sort (supersymmetry, technicolor or something else) around the TeV mass scale is required to explain the spontaneous breaking of electro-weak symmetry without introducing the hierarchy problem. The hope that new colliders will probe this new physics and produce new particles $\chi$ associated with it is a main driving force in high energy physics.

For a given mass $M_\chi$ in the TeV range, the $\chi-\bar{\chi}$ production cross-section in the hadronic LHC collider is much larger when $\chi$ carries color rather than just $SU(2)\times~U(1)$ quantum numbers. If, however, these massive colored particles (MCP's) are as light as the other new particles, than they can be long lived or even stable. A concrete interesting example, in which this might happen, is the split supersymmetry scenario with gauginos and gluinos in particular much lighter than the squarks \cite{Arkani-Hamed:2004fb}.

For reheat temperatures exceeding the mass $M_\chi$, the cosmological abundance of the particles remaining after freezout at temperatures $T \sim M_\chi/30$ is rather small \cite{Kang:2006yd}:
\begin{equation}\label{eq:abundance}
f_\chi= \frac{n_\chi}{n_{\gamma}} \sim 10^{-14},
\end{equation}
where $n_\gamma$ is the entropy density at freezout. This is small enough to prevent such particles from playing any role as dark matter often assumed to arise in a similar fashion via lightest supersymmetric partners annihilating with considerably weaker annihilation cross-sections. Late decays of even such a tiny number of particles can, however, lead to excessive (and observed) fluxes of high energy gamma rays, modifications of BBN (big bang nucleosynthesis) and/or distortion of the CMBR (cosmic microwave background radiation).

The $\chi-\bar{\chi}$ annihilation cross-section is a key ingredient in relic abundance estimates. The attractive colored gluon exchange interaction enhances the cross-section. Yet, since $\chi-\bar{\chi}$ strongly annihilate in $S$-wave only, unitarity implies the general upper bound on the annihilation cross-section \cite{Griest:1989wd} (which is $\mathcal{O}({{\alpha}_{s}}^{-2})$ larger than the perturbative value):
\begin{equation}
\frac{4\pi}{M^2_\chi \beta} \sim \frac{1}{M_\chi T} > \sigma(\chi \bar{\chi})_ {annihilation},
\end{equation}
which again leads to excessive relic abundance of $\chi$'s.

It has been noted \cite{Arkani-Hamed:2004fb}, however, that after the QCD confining phase transition at temperature $T_c \sim \Lambda_{QCD} \sim 200$ MeV, relic MCP's combine with quarks or gluons making $\bar{Q}q$ mesons or $g\tilde{g}$ glueballinos (``$Q$'' $\chi$'s are
$SU(3)$ triplets and ``$\tilde{g}$'' $\chi$'s are $SU(3)$ octets). The scattering cross-section of these hadrons on each other is geometric and rather large:
\begin{equation}
\sigma_{scattering} \sim {\pi} R_H^2
\end{equation}
where $R_H=1/{\Lambda}_H$ is the hadrons' radius.

The very large mass $M_\chi \sim$ TeV of the MCP's, and corresponding large momentum at a given energy (temperature) implies that many partial waves are involved in the collision (which is indeed required in order to avoid the $S$-wave unitarity upper bound). Thus, for $T \sim T_c \sim 200$ MeV the scattering involves partial waves up to
\begin{equation}
L \sim p\times R_H \sim \sqrt{2M_\chi T}\times R_H \sim 20,
\end{equation}
for the small (conservative)
\begin{equation}
    R_H =1/{{\Lambda}_H} =(\mathrm{GeV})^{-1}=0.2\ \mathrm{Fermi}
\end{equation}
used in \cite{Kang:2006yd}, and $M_\chi=1$ TeV. This relaxes the above unitarity bound by a factor of $L^2 \sim 400$.

The large hadronic heavy meson-heavy meson cross-sections enhances the $\chi\bar{\chi}$ annihilation cross-section and reduces the relic abundance of the $\chi$'s in eq. (\ref{eq:abundance}) by about three more orders of magnitude to $f_\chi \sim 10^{-17}$, thereby resuscitating most MCP scenarios even when the colored massive particles decay rather late \cite{Kang:2006yd}. It is difficult to quantitatively calculate the rate of these late annihilations and in the following we further discuss this problem.

First we address two simple issues:
\begin{itemize}
\item In a large fraction of collisions of heavy meson and anti-meson the latter rearrange into a tightly bound heavy-heavy $Q\bar{Q}$ Quarkonium or $\tilde{g}\tilde{g}$ and a light $q\bar{q}$ meson or a $gg$ glue-ball:
    \begin{eqnarray}
        &Q\bar{q}&\!\!+\bar{Q}q \rightarrow Q\bar{Q}+q\bar{q} \nonumber \\
        &\mathrm{or}& \nonumber \\
        &\tilde{g}g&\!\!+\tilde{g}g \rightarrow \tilde{g}\tilde{g}+gg.
    \end{eqnarray}

  The anomalous lightness of the pion and the parametrically large binding in the heavy-heavy system (the binding energy increases with $M_\chi$) ensure an exothermic process even for excited $\chi\bar{\chi}$ final states. Since the rearrangement involves transition to a different ``branch'' it is not adiabatically suppressed. The transition can happen at any point during the long collision time of the heavy $\chi$ hadrons:
    \begin{equation}
      t_{collision} \sim \frac{R_H}{\beta_\chi}
    \end{equation}
    with
    \begin{equation}
        \beta_\chi \sim \sqrt{\frac{2T}{M_\chi}} \sim 0.02
    \end{equation}
and its probability should indeed be $\sim 1$ as argued in Ref. \cite{Kang:2006yd}.
 We note, however, that for quark-like $\chi$'s the relevant rearrangement collisions may involve a heavy baryon and a heavy meson:
    \begin{equation} \label{eq:nucleon}
        Qqq+\bar{Q}q \rightarrow Q\bar{Q}+qqq \ \textrm{(Quarkonium + nucleon)}
    \end{equation}
 rather than two heavy mesons as assumed in Ref. \cite{Kang:2006yd}:
    \begin{equation} \label{eq:pion}
      Q\bar {q}+\bar{Q}q \rightarrow Q\bar{Q}+\bar{q}q \ \textrm{(Quarkonium + meson)}
    \end{equation}
  The reason is the following. At the time of these hadronic assisted late $\chi$ annihilations (at and somewhat after the QCD phase transition), the baryon to entropy ratio $\sim 6\cdot 10^{-10}$ vastly exceeds the $\chi$ to entropy ratio $\sim 10^{-14}$. Hence, a $Q\bar q$ heavy meson is $4 \cdot 10^4$ times more likely to collide first with a nucleon and transform into a heavy baryon\footnote{For reaction \ref{eq:baryon_formation} to proceed, even for low kinetic energies of the colliding hadrons, the sum of their masses should exceed the sum of masses of the final two hadrons. This is guaranteed by the light final pion and the binding of the $ud$ $I=S=0$ diquark to massive $Q'$ which is stronger than the binding to a light $u$ as in the proton. Indeed, for $c$ and $b$ quarks, the corresponding differences are already positive and large:
  $m_D+ m_p -  [m_{\Lambda_c}+m_\pi]=380$~MeV and  $m_B +m_p - [m_{\Lambda_b}+m_\pi]=450$ MeV.}:
  \begin{equation}\label{eq:baryon_formation}
    Q\bar{q}+qqq \rightarrow Qqq+\bar{q}q \ \textrm{(Heavy baryon + meson)}
  \end{equation}
 rather than collide directly with the rare heavy anti-meson. The baryon will eventually annihilate via eq. (\ref{eq:nucleon}).

 The lighter pion emitted in eq. (\ref{eq:pion}) allows more loosely bound and larger Quarkonia to form than in the case of reaction (\ref{eq:nucleon}). The resulting, slightly reduced, cross-section of reaction (\ref{eq:nucleon}) (as compared with reaction (\ref{eq:pion})) and ensuing decrease of late annihilation rate are likely to be moderate and not effect the qualitative results of Ref. \cite{Kang:2006yd}.
\item Most of the collisions and quarkonia formed therein have high orbital angular momentum\footnote{This feature and ensuing geometric cross-section exclude the extra $1/\beta$ enhancement of $S$-wave exothermic processes}. The centrifugal barrier quenches $\chi\bar{\chi}$ annihilations in these states and cascading to the ground $S$-wave state needs to be investigated.
\end{itemize}

The authors of Ref. \cite{Kang:2006yd} assume that these states lie within the $\sim 1$ GeV deep linear regime of the potential between the Heavy quarks with binding energy
\begin{equation}
    B.E \arrowvert_{_{initial}} \lesssim 400\ \mathrm{MeV}.
\end{equation}

Since there is no $\chi\bar{\chi}$ annihilation in these high angular momentum $\chi\bar{\chi}$ states the rate of cascading into lower energy and lower angular momentum states is indeed relevant.

If the $\chi$'s were electrically charged, cascading down from an initial, relatively loosely bound state in the linear regime of the iterquark potential,
\begin{equation}
    V(r)=\frac{C \cdot \alpha_{QCD}}{r}-\sigma \cdot r,
\end{equation}
to the more tightly bound $\chi\bar{\chi}$ states in the QCD Coulombic regime lasts a very short time \cite{Kang:2006yd}:
\begin{equation}
    t_{cascade} \sim \frac{\sqrt{\alpha_{QCD}} \cdot M_\chi^2}{\alpha_{em} \Lambda_H^3} \sim 10^{-16} \ \mathrm{sec}.
\end{equation}

Possible lack of anomaly cancelations and/or large $SU(2)_L \times U(1)$ breaking by massive fermions which are not neutral under these groups suggest that the fermions are electrically neutral.

The authors of Ref. \cite{Kang:2006yd} were overly conservative in estimating the cascading time for neutral MCP's. Assuming that the cascade proceeds only via two photon emission at each stage and using effective lagrangian/dimensional arguments they find a long cascading time of $\mathcal{O}(1)$ sec.

Even for neutral $\chi$'s, however, one photon $\Delta L=1$ transition between quarkonium states of opposite $C$ are allowed. The photon converts via a light quark loop into a $C= -1$ color singlet three gluon state, which couples to the heavy quarks (see fig. \ref{fig:photon}).
\FIGURE{\label{fig:photon}
\begin{fmffile}{photon}
\begin{fmfgraph*}(240,170)
    \fmfleft{i1} \fmfright{o1} \fmftop{o2}
    \fmflabel{$L=1$}{i1}
    \fmflabel{$L=0$}{o1}
    \fmf{plain,tension=100,width=3}{i1,v1,v2,v3,o1}
    \fmf{gluon,tension=0.8}{v1,v4}
    \fmf{gluon,tension=0}{v2,v5}
    \fmf{gluon,tension=0.8}{v3,v6}
    \fmfpoly{empty,smooth,tension=1}{v4,v5,v6,v7}
    \fmf{photon,label=$\gamma$,tension=2}{v7,o2}
 \end{fmfgraph*}
\end{fmffile}
\caption{Emission of a single photon from the process $B'
\rightarrow B'\gamma$. The thick line is the bound state of the heavy colored particles and the
thin lines are light quarks.}}

Following Ref. \cite{Kang:2006yd} the corresponding effective lagrangian is now:
\begin{equation}
    \mathcal{L} \sim \frac{e}{\Lambda} F^{0,j} \bar{\psi}\frac{i\partial_j}{M_\chi} \psi,
\end{equation}
with a decay width:
\begin{equation}
    \Gamma(B'\rightarrow B \gamma) \sim \frac{\alpha E_\gamma^3}{M_\chi^2{\Lambda_H^2 r_B^2}},
\end{equation}
where $r_B$ is the size of the bound state. This rate\footnote{This is a very crude estimate: The two (rather than three) body phase enhances the one photon process. On the other hand, the partial cancelation due to $q_d+q_u+q_s=0$ between the three diagrams with the three light quark loops in fig. \ref{fig:photon} suppresses it relative to the two photon cascade.} is
\begin{equation}
   \sim 137 \left(\frac{\Lambda_H}{E_\gamma}\right)^2 \sim 10^6 -10^7
\end{equation}
times faster than the two photon rate of Ref. \cite{Kang:2006yd} for $E_{\gamma}\sim {\Delta}E \sim 100-50$ MeV and as in Ref. \cite{Kang:2006yd} $\Lambda_H \sim$ GeV.

This dramatic enhancement of the rate of cascading in the quarkonia system supports the argument that ``late'' annihilation of MCP's after the QCD confinement will reduce the abundance of the latter so as to meet all the astrophysical bounds.

We note also that hadronic radius
\begin{equation}\label{eq:R_H}
    R_H \sim \frac{1}{2}\ \mathrm{Fermi}
\end{equation}
is appropriate for heavy meson-heavy meson case. Using this, rather than the conservatively (small) value $R_H \sim \mathrm{GeV}^{-1} \sim 0.2\ $Fermi used by Ref. \cite{Kang:2006yd}, enhances the annihilation cross-section (and reduces the expected final CMP's density by $\sim 6$).

So far, our comments tended to enhanced the CMP's annihilation supporting the conclusions of \cite{Kang:2006yd}. The following appears to have the opposite effect.

Ref. \cite{Kang:2006yd} assumes that the cosmic background pions have no impact on $Q'\bar{Q'}$ bound state formation which happens at a temperatures $T \sim T_B$. For the temperature $T_B$, however, used for estimating the final relic density:
\begin{equation}
T_B \sim T_c \sim \Lambda_{QCD} \sim 180\ \mathrm{MeV},
\end{equation}
this assumption is not justified. At this temperature the pions' density is comparable to that of photons' and larger than that of the heavy $\chi$ particles surviving the earlier perturbative annihilation phase after freezout at $T \sim M_\chi/{30}$ by $10^{14}-10^{17}$, corresponding respectively to the beginning and end of the late annihilation phase.

In passing we note that despite the electromagnetic and weak decays of $\pi^0$ and $\pi^+$ which are much faster than the Hubble expansion at this time, the decay products (muons, electrons, photons and neutrinos) are in chemical equilibrium with the pions, immediately reinstating the latter at their equilibrium value.

The inverse of the reaction in eq. (\ref{eq:pion}) above, where the Quarkonium absorbs a pion with energy exceeding its binding,
\begin{equation}
   E \geq |B.E|=2 M_{Q'\bar{q}}-M_{Q'\bar{Q'}},
\end{equation}
destroys the newly formed $\chi \bar{\chi}$ state on a typical hadronic time scale
\begin{equation}
    t \sim \frac{1}{T_B} \sim 10^{-24}\ \mathrm{sec},
\end{equation}
vastly shorter than the time for any of the ``autonomous'' cascading down considered so far. This stops only when the number density of pions with the requisite energy (which decreases with a phase space and Boltzman weight factor $E_\pi^{3/2}e^{-E_\pi/T}$) becomes comparable to that of the $\chi$ particles at temperatures $\sim E/30$ when the exponential factor overcomes the huge initial $10^{14}$ ratio. Hence, only at temperatures $\sim 1/30$ of the binding energy of the heavy quarkonia state initially formed, will most quarkonia become immune to break-up via pion absorbtion and start cascading to lower states.

This is analogous to the well studied, ordinary electron-proton recombination into hydrogen which occurs only at temperatures of
\begin{equation}
    T_{recombination} \sim \frac {1}{40}Ry \sim \frac{13.6}{40} \ \mathrm{eV} \sim 0.3 \ \mathrm{eV}
\end{equation}
due to the large number of photons which can break up the bound hydrogen (see also \cite{Wolfram:1978gp} in modern context).

The suggested value of the temperature when the bound states effectively form is then:
 \begin{equation}
    T\sim \frac{B.E}{30} \sim \frac{400}{30}\ \mathrm{MeV} \sim 13 \ \mathrm{MeV},
 \end{equation}
namely $14$ times lower than the temperature $T_B \sim 180$ MeV used in Ref. \cite{Kang:2006yd}. This increases the residual relic abundance of $\chi$'s which is given in Ref. \cite{Kang:2006yd} by
\begin{equation}
Y_\chi=\frac{n_\chi}{s} \sim 10^{-17} \left(\frac{R}{\mathrm{GeV}^{-1}}\right)^{-2} \left(\frac{T_B}{180\ \mathrm{MeV}}\right)^{-3/2} \left(\frac{m}{\mathrm{TeV}}\right)^{1/2},
\end{equation}
by a factor of $\sim \left(1/14\right)^{-3/2} \sim 50$.

There is yet another crucial factor, however, that should be taken into consideration. Collisions with ambient pions will not only break the newly formed $Q'\bar{Q'}$ states (once $E_{\pi} > B.E$). The pion can also scatter off the quarkonium state leaving the latter more tightly bound and with angular momentum $L-2$ instead of $L$. The emitted pion then carries the two units of angular momentum and its energy is increased by the binding energy difference
\begin{equation}
\Delta E=B.E(L-2)-B.E(L).
\end{equation}

We can estimate the actual cross-section for this pion induced downward cascade, which is much faster than any autonomous cascade mentioned earlier.
Unlike the case of pion absorbtion, where the quarks from the pion are incorporated into the two $Q'\bar{q}$ mesons, the process at hand requires two gluon exchange. This process is analogous to inelastic diffraction, which is well known from ordinary hadron high energy scattering. The difference is that there, the colliding particles are in the ground state and one or both become excited, whereas here, the $Q'\bar{Q}'$ target is a highly excited state and we are interested in the de-excitation to a lower state. The (induced) color dipole-color dipole gluon exchange interactions explain the well known geometric nature of such cross-sections \cite{Nussinov:1975mw}. Both the pion and the initial highly excited $Q'\bar{Q}'$ state have normal hadronic size. Hence, we expect roughly equal $\sigma_{breakup}$ - the cross-section for breakup via pion absorption and $\sigma_{de-excitation}$, the cross-section for diffractive de-excitation. Diffraction selection rules and the limited angular momentum carried by the light pions imply that only levels with angular momenta lower by two units are likely to be reached in the de-excitation reaction. This and the extra powers of $\alpha_s$ involved suggest that the de-excitation cross-section for the downward induced cascade is smaller than that of the quarkonium breakup by:
\begin{equation}
\left(\frac{\alpha_s E_{\pi}}{\Lambda_H}\right)^2 \sim 10^{-2} -10^{-3},
\end{equation}
were $E_\pi \sim m_\pi$ is the energy of the absorbed pion.

A much larger number (by a factor $F \sim e^{(B.E+m_\pi)/T}$) of pions, however, can generate a downward cascade as compared with the number of those that are energetic enough to break the initial bound quarkonium state.

Once the temperature drops bellow $\sim 100$ MeV, so that this factor exceeds $\sim 10^2-10^3$ and compensates for the larger break-up cross-section, the quarkonium system will cascade down to ever more tightly bound and harder to break states (and eventually annihilates) before meeting a ``killer'' pion of sufficiently high energy which breaks it up. This will happen for temperatures $T_B$ larger than $B.E/30$ used above, and much closer to $T_B~ \sim~180$~MeV used by Ref. \cite{Kang:2006yd}. Thus, the dangerous factor of $\sim 50$ above largely disappears and
the final conclusion is, thus, that the further $\mathcal{O}(10^{-3})$ reduction of the relic abundance of putative massive colored particles via enhanced annihilations after the QCD phase transition and hadron formation found by the authors of Ref. \cite{Kang:2006yd} will be reinstated.

The above discussion utilizes the known pattern of masses of heavy quarkonia and of mesons involving one heavy $b$ or $c$ quark extrapolated to even heavier quarks. If the MCP's are gluinos rather than heavy quarks, the systematics of the transitions of
\begin{equation}
    \tilde{g}g+\tilde{g}g \rightarrow \tilde{g}\tilde{g} + gg
\end{equation}
may be quite different. In particular there is no anomalously light glueball, which is the analog of the $q\bar{q}$ pion. Thus, the gluino relic abundance will be reduced to a lesser extent after the QCD phase transition and may have less impact. A single pion emission is forbidden here by isospin and two pion emission will be suppressed by extra powers of $\alpha_s$. Still, the heavy gluinos with larger $SU(3)$ Casimir will be very
tightly bound and transitions of this type with generic hadronic cross-section are expected.

\section{Can new confining gauge theories manifest via macroscopic large strings?}
\subsection{Introduction}

String theory originated in the 1960's in an effort to explain the hadronic spectrum, which presently is being explained by the local gauge theory of QCD.

The chromo-electric flux tubes, which are believed to connect color charges generating a confining linear potential, are the vestige of such strings. When the tube/string gets longer than some critical distance of order $1/\Lambda_{QCD}$, however, a Schwinger pair creation mechanism of light ($u$, $d$ and possibly $s$) quark anti-quark causes it to break. Only when the lightest quarks carrying the fundamental representation $3_c$ are much heavier than $\Lambda_{QCD}$ can the strings live long enough \cite{Gupta:1981ve}. If excited to high enough energy, the mesons get elongate, so that they look like strings rather than almost spherical bags, and keep oscillating until slowly dissipating via glueball emission.

The following discussion is largely inspired by the suggestion of Markus Luty, (Unpublished) of the ``Quirk model'', a model in which macroscopic strings may arise. Within the frame-work of such models we address the question: Is there a consistent field theory and cosmological scenario where macroscopic strings arise?

Luty's Quirk model seems ad hock and does not address any problems in the standard model or extensions thereof. The extra groups, fermions and confining flux tube do not ease the hierarchy problem, do not explain dark matter and/or dark energy and/or any other possible astrophysical anomaly. Still we find the possible existence of macroscopic strings which, unlike the cosmic strings, can actually be manipulated, so fascinating to justify the following discussion.

We will begin by briefly reviewing the model in section \ref{subsec:introduction}. We next discuss in section \ref{subsec:cosmological} the cosmological implications of the new $SU(N')$ gluons and the $SU(N')$ glueballs and ensuing limits on the model. A rather detailed estimate of the relic abundance of the Quirks surviving in this model to the present date follows in \ref{subsec:estimate}, with special further discussion of baryonic $Q'^{N'}$ type states in section \ref{subsec:baryonic}. These two sections extend the discussion in section \ref{sec:reduced} above on ``hadron assisted late annihilations'' after the $SU(3)_c$ confinement to account for ``string assisted very late annihilations'' occurring after the $SU(N')$ confining phase transition. Section \ref{subsec:LHC} briefly comments on the possible manifestation of the new long confining strings in LHC. In section \ref{subsec:fate} we follow the evolution the $Q'\bar{Q'}$ pairs with attached strings produced at LHC. We find that for small $\Lambda'$s it is quite probable for the ends of one $SU(N')$ string to be trapped in separate chunks of matter. Finally in section \ref{subsec:trapped} we speculate on the fascinating physics that would ensue if the above scenario is indeed realized and the separate string ends can be manipulated enabling us to verify the existence of such macroscopic strings and directly measuring the string tension.

\subsection{Introduction of the Model} \label{subsec:introduction}

The following describes Luty's Quirk model. Unfortunately, his long anticipated paper on this subject has not come out yet so the following should be considered as a rough sketch. It does, however, suffice for our main goal, namely discussion of the cosmology and some phenomenology.
The standard $SU(3)_c \times SU(2)_L\times U(1)_Y$ with three fermion generations of quarks and leptons is extremely successful. Yet, there is no true understanding why the above specific groups and (triplicated) fermionic representations are chosen. In particular there is no true understanding of the mass scales in QCD and in the electroweak sector and of fermion masses. The $u$ and $d$ quarks are $10-100$ times lighter than the QCD scale, yet, a-priori we could have only quarks (for instance third generation quarks) with masses much larger than $\Lambda_{QCD}\sim 200$ MeV.

In view of the above, the following modification of the standard model may not be too unnatural: An extra $SU(N')$ non-abelian gauge group is added as a direct product factor to the standard model. Just like QCD, it is assumed to be vectorial and confine at some scale $\Lambda'$. Furthermore, there are fermions (``Quirks'') denoted by $Q'$, which transform as its fundamental $N'$ representation. Next, in analogy with having the ordinary quarks transform not only under $SU(3)_c$ but also under the other product groups $SU(2)_L$ and/or $U(1)$, the quirks are assumed to also carry a $3_c$ representation\footnote{A priori we could have $(N',\bar{3}_c)$ fermions. Either choice will lead to the same conclusions.} of ordinary vectorial $SU(3)_c$. The simplest way to maintain the cancelations of axial anomalies and avoid excessive breaking of $SU(2)_L \times U(1)$ by the mass of the heavy $Q'$s is to keep the quirks $SU(2)_L \times U(1)$ neutral.

The expression for the QCD $\beta$ function becomes:
\begin{equation}
    \beta(g)=-\frac{g^3}{(4\pi)^2}\left(\frac{11}{3}N_c-\frac{2}{3}n_f-\frac{2}{3}N'\right)
\end{equation}
where $N_c=3$, $n_f=6$, and we assume the new $SU(N')$ group has a single flavor. Thus, $N'<11$ is required to maintain QCD asymptotic freedom. As we will see below, cosmology implies a much more severe upper bound of $N'\leq 3$.

Since the ordinary and the new $SU(N')$ color are conserved, the $Q'$s which are the lightest particles carrying both are absolutely stable. We need therefore, to extend the discussion in section \ref{sec:reduced} above and to verify that $Q'-\bar{Q}'$ annihilations at various stages reduces their relic abundance so as to meet all bounds. The general discussion for any massive colored particle was given above and further effects of $SU(N')$ interactions will be discussed below.

To explain why quirks have not been produced to date in colliders, their masses should satisfy $M_{Q'} \geq 300$ GeV and to allow production at the new hadron collider LHC we assume $M_{Q'} \leq$TeV. We shall use $1$ TeV as the nominal mass in the following.

The key to the new fascinating phenomenology is the assumed extreme smallness of the $SU(N')$ scale:
\begin{equation}
\Lambda' \sim 10- 10^5 \ eV.
\end{equation}

While all $\Lambda'$s in this range lead to small cosmological relic abundances a $10-100$ eV value is particularly suggestive. Such $\Lambda'$s may allows the atoms of the nuclei to which the $Q'$s at the ends of the $SU(N')$ string attach, to remain within the crystals and yet have measurable string tension of $10^{-5}-10^{-4}$ dyne.

This choice seems to require extreme fine-tuning. This, however, is not the case. The scale of a non-abelian gauge theory confinement is the mass scale at which the running coupling constant becomes of order unity. Because of the logarithmic variation,
\begin{equation}
\alpha' \sim \frac{\beta'_0}{ln{Q^2}},
\end{equation}
the scale is exponentially sensitive to the value of the coupling at some standard energy, $1$~GeV for example, and on the number of colors/flavors which determine the beta function.

Thus, the scale could readily be $10^3$ times larger than the ordinary QCD scale (as was assumed in technicolor theories, designed to explain the $SU(2) \times U(1)$ breaking scale), or $10^{-6}-10^{-7}$ smaller as is assumed here for $SU(N')$.

\subsection{Cosmological Implications}\label{subsec:cosmological}
The introduction of the extra gauged $SU(N')$ modifies the various stages in the more general cosmological scenario with massive colored particles discussed above in several ways. We will discuss those next starting with the earlier and continuing with later stages.

\begin{itemize}
\item The abundance of the MCP's, the stable $Q'$s in the present case, remaining after the
color assisted late annihilations is hardly effected by the new $SU(N')$. We have, in addition to the $Q'\bar{Q}'$ annihilation into ordinary gluons (or quarks), also annihilations into the ${N'}^2-1$ gluons . This will slightly reduce the $Q'$ abundance (of $n_{Q'}/s \sim 10^{-14}$) after the early annihilation stage freezes out at $T \sim m_{Q'}/30$. The annihilation cross-sections, however, are proportional to the squares of the corresponding $\alpha$'s and with $\Lambda_{QCD} >> \Lambda'$, $\alpha$' is much smaller than $\alpha_s$ at energies $\sim m_{Q'}$ and this effect is minimal.

\item The further reduction by another factor of $10^3$ of $Q$' abundance after QCD confinement and formation of heavy $Q'\bar{q}$ hadrons, crucial for allowing long lived MCP's, was discussed at length above. The arguments leading to this reduction are strengthened by the additional $SU(N')$ interactions which accelerate the cascade to the lower more strongly bound $Q'\bar{Q}'$ states. The point is very simple. The $SU(N')$ interactions become strong only at a scale of order $10-100$ eV or equivalently distances of order $20-200 \ \AA$, however, the size of the $Q'\bar{Q}'$ states which form first $R_H \sim 1/2$ Fermi is almost $10^7$ smaller.
At such ``short distances'' $SU(N')$ is still perturbative, and like for ordinary gluon jets for multi TeV QCD processes we need not worry at all about the effect of eventual $SU(N')$ confinement. Thus, even for electrically neutral $Q'$s, we have the one $g'$ de-excitation mechanism
    \begin{equation}
        Q'\bar{Q}'|_L \rightarrow Q'\bar{Q}'|_{L-1} + g'.
    \end{equation}

    Furthermore, since the coupling constant
    \begin{equation}
        \alpha'(\Lambda') \sim \mathcal{O}(1),
    \end{equation}
    and decreases logarithmically, we expect that at the heavy quarkonium scale
    \begin{equation}
      1 \gg \alpha' > \alpha_{em} \sim 1/137.
    \end{equation}
    This makes then the $g'$ emission a most efficient ``autonomous'' de-excitation mechanism and following Ref. \cite{Kang:2006yd} and the discussion above the complete cascade will terminate on very short time scales of $10^{-17}-10^{-16}$ sec.

\item The ${N'}^2-1$ $g'$ gluons of the new gauge group exist as radiation in the primordial plasma at the time of nucleosynthesis and at temperatures
    \begin{equation}
        T \sim MeV \gg \Lambda'.
    \end{equation}

 The smallest non-abelian groups $SU(N')$ with $N'=2\ (3)$ have three (eight) new $g'$s. Up to a factor of $7/8$, each massless vector gluon is equivalent in terms of number of statistical degrees of freedom to a Majorana neutrino. The success of big bang nucleosynthesis calculations limits the number of relativistic degrees of freedom, and the number of neutrino species at that time to $N_{\nu} \leq 3\pm \Delta$, where $\Delta \lesssim 1$. It seems, therefore, impossible at first sight to reconcile the success of big bang nucleosynthesis calculations limiting the total number of degrees of freedom at that time and with the existence of three (eight) extra $g'$s in thermal equilibrium. As we show next, however, this scenario \emph{is} consistent with $N'^2-1$ extra $g'$s for $N'<4$, that is up to \emph{eight} extra gluons equilibrium.

    The $g'$ and ordinary gluon and quarks couple only via the very heavy $Q'$s and completely decouple after the first $Q'$ annihilation stage one freezout at
    \begin{equation}
        T \sim \frac{M_{Q'}}{30} \sim 30 \ \mathrm{GeV}.
    \end{equation}
    In fact, since $\alpha'$ is likely to be significantly weaker at such energies than $\alpha_s~\sim~0.15$ the $g$'s may decouple even earlier .

 The photons, electron-positron pairs and neutrinos at $T \sim$ MeV have, however, been enriched by the annihilation of all degrees of freedom in the standard model's $16$ gluons, $(7/8) \cdot 3 \cdot 28 \sim 74$ degrees of freedom due to the (non neutrino) fermions in the three generations and the $3 \cdot 4 + 1=13$ degrees of freedom due to the massive $W^+,\,W^-,\,Z^0$ and Higgs. Adding to this the $2+6 \cdot 7/8$ degrees of freedom in the photons and neutrinos themselves this sector has altogether $110$ degrees of freedom of which $1/2$ resides in the neutrinos and a $1/6$ namely $\sim 18.5$ from each neutrino. The $6-16$ degrees of freedom of the $g'$s are equivalent to $\Delta \sim 1/3-1 $ additional neutrinos and is cannot be excluded.

 The above discussion is hardly affected when we take into account the fact that the $6-16 \cdot 7/8=5-25$ degrees of freedom in the $Q'$s do also leak in part into the $g'$s. Since, however, the $Q'$s couple to all colored degrees of freedom of $8$ gluons and $6$ colored quarks (equivalent to $89$ degrees of freedom and only six degrees of freedom of $g$'s) even a purely statistical division would increase the degrees of freedom in $g$'s by only $\sim 1/16-1/6 $, and even that is an upper bound since the $g$'s may decouple earlier.

\item Once the temperature drops bellow
    \begin{equation}
        T'_C \sim \Lambda' \sim 10-100 \ \mathrm{eV},
    \end{equation}
    all $g'$s combine into $SU(N')$ singlet glueballs. Lattice QCD in its simplest ``quenched'' form (which is completely justified here) implies for $N'=3$ that the lightest glueball is a $0^{++}$ particle of mass $\sim 7\Lambda'$ \cite{Chen:1993tr,Vaccarino:1999ku}. The glueballs can decay into photon pairs via $Q'$ loops. The rate of decay, however, (even in the most favorable case where the $Q'$s carry electric charge and no further $SU(3)_c$ gluons and light quark loops are needed) is negligible:
    \begin{equation}
        \Gamma \sim \frac{{\alpha'}^2 \alpha_{em}^2 \cdot {\Lambda'}^9}{{M_{Q'}^8}} \lesssim 10^{-70} sec^{-1}
    \end{equation}

    Thus, the scalar glueball's are practically stable warm dark matter, which, for the smallest considered scale, $\Lambda' \sim 10$ eV, might even saturate dark matter. This poses a bit of a problem, as cold dark matter is preferred. The scalar glueballs constitute a particularly interesting form of ``warm'' dark matter, in which number changing reactions of the form
    \begin{equation}
        3 gb's \rightarrow 2 gb's
    \end{equation}
    can occur. As the universe expands, the glueballs adiabatically cool and the inverse reaction eventually stops. The co-moving number of the glueballs is no longer conserved and keeps decreasing for some time with a Boltzman factor $e^{-m_{gb'}/T}$ . The process freezes out at $T' \sim m_{gb'}/{30}$ and, as shown in some detail by Ref. \cite{Carlson:1992fn}, $n_{gb'}/s$ becomes extremely small.
\end{itemize}

\subsection{Estimate of the Relic Abundance of the Quirks} \label{subsec:estimate}
As noted in section \ref{sec:reduced} above (see also Ref \cite{Raby:2007hm}), the $Q'$s make $Q'ud$ baryons (rather than $Q'\bar{q}$ mesons) and the anti-Quirks make $\bar{Q'}u$ mesons\footnote{The $\bar{Q'} d$ member of the isospin doublet is $(m_d-m_u) \sim 3\ \mathrm{MeV}$ heavier and Beta decays in $ \mathcal{O}(\mathrm{sec})$ into this lighter one. One second is the analogous charged to neutral pion beta decay partial lifetime}. These heavy- light baryons and mesons carry (\emph{fractional}) charges, have strong hadron like interactions and will bind with heavy nuclei to form \emph{anomalous} heavy isotopes.

Thus, the very stringent bounds on such isotopes and/or fractional charged particles apply, limiting the $Q'$ terrestrial number density to be very small:
\begin{equation}
    \frac{n_{Q'}}{n_B} < 10^{-30}.
\end{equation}
Accounting for possible extra concentration in galaxies, in the solar system and in earth we may need to limit the $Q'$ to entropy ratio to be even smaller than $<10^{-40}$, far below the $\sim 10^{-17}$ value achieved after the hadronic assisted annihilation stage described in section \ref{sec:reduced} above.

The $SU(N')$ permanent confinement of the $Q'$s sets in at a temperature
\begin{equation}
    T'_C \sim \Lambda' \sim 10-10^5\ \mathrm{eV}
\end{equation}
is expected to bring all $Q'\bar{Q'}$ relics surviving till this stage, close together, leading eventually to annihilation. Thus barring appreciable $Q'$ asymmetry\footnote{Gauge coupling unification is lost in the simple variant. Furthermore, the $Q'$s carry no chiral charges and there is no 't Hooft $U(1)$ anomaly \cite{'tHooft:1980xb} for the axial ``$Q'$ baryon number'', nor are there new leptons associated with the Quirks. Thus two mechanisms, which can account for the ordinary baryon asymmetry, are absent here, and the Quirk asymmetry, while possible, seems unlikely.} (analogous to the ordinary baryon asymmetry) we expect practically complete annihilation of all $Q' \bar{Q'}$ and no ``dangerous'' present relic concentration. In this section and in the following section we show that this expectation is indeed born out. Further, all annihilation occur early enough and any imprint of annihilation products is washed away.

In passing we note, that the entities initially confined by $SU(N')$ are not the $Q'$s but rather the above hadrons composed of $Q'$s and light ordinary quarks. The $\mathcal{O}(\mathrm{Fermi})$ size of these hadrons is far smaller than $\Lambda'^{-1}$, the $SU(N')$ confinement scale. Thus, we can view them as point-like objects in the $N'$ representation for most of the evolution, until the very last stages when the light $u$ and $d$ quarks get stripped from the $Q'$ (and $\bar{Q'}$) with the latter rearranging into a $Q'\bar{Q'}$ Quirkonium state, bound via ordinary color $SU(3)_c$ forces.

The average separation between photons or between the $SU(N')$ glueballs $gb$'s at $T \sim T' \sim \Lambda'$ for $\Lambda'$ in the $10$ eV to $10^5$ eV range is:
\begin{equation}
     d_{gb',gb'}\sim \frac{1}{T'} \sim 0.002-200 \AA.
\end{equation}
The density of the $Q'$s at this stage is
\begin{equation}
    n_{Q'} \lesssim 10^{-16} n_{gb'},
\end{equation}
so that the average $Q'Q'$ (or $Q'\bar{Q'}$) separation is
\begin{equation}
    f \sim 2\cdot 10^5
\end{equation}
times larger.

The setting in of $SU(N')$ confinement means that flux tubes/strings connecting nearby $Q'$ and $\bar{Q'}$ form. The light $gb$'s and the string with small tension constitutes fast degrees of freedom and the heavy $Q'$s at the strings ends move slowly. This suggests a Born-Oppenheimer approximation, where at any time the system achieves the total minimal length string network connecting the heavy $Q'$s and $\bar{Q'}$s.

The initial string length which is on average $L_1 \sim d_{Q'Q'}$, exceeds by the same large factor $f \sim 2.10^5$ the expected size of Quirkonia in thermal equilibrium, namely
\begin{equation}
   R' \sim \frac{T'}{\sigma'} \sim \frac{1}{\Lambda'} \sim 2-20 \cdot 10^{-7}\ \mathrm{cm},
\end{equation}
where $\sigma' \sim \Lambda'^2$ is the string tension and $T' \sim T'_C \sim \Lambda'$.

Two different processes can dissipate the energy of the initial, long strings and ``relax'' them into the final small size $R'\sim \Lambda'^{-1}$ . These involve:

\begin{enumerate}
\item Interactions of a single $Q'Q'$ string with the thermal bath of $SU(N')$ $gb$'s.
\item String string scattering leading in $\sim 50\%$ of the cases to string reconnection, followed by straightening and shortening of the resulting new bent strings (fig. \ref{fig:strings}).
\end{enumerate}

\FIGURE{\label{fig:strings}
\begin{picture}(0,0)%
\includegraphics{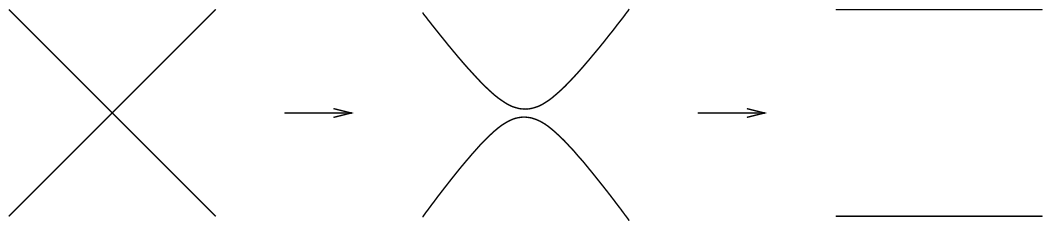}%
\end{picture}%
\setlength{\unitlength}{2763sp}%
\begingroup\makeatletter\ifx\SetFigFont\undefined%
\gdef\SetFigFont#1#2#3#4#5{%
  \reset@font\fontsize{#1}{#2pt}%
  \fontfamily{#3}\fontseries{#4}\fontshape{#5}%
  \selectfont}%
\fi\endgroup%
\begin{picture}(7307,1866)(1214,-1767)
\put(4064,-1689){\makebox(0,0)[lb]{\smash{{\SetFigFont{8}{9.6}{\rmdefault}{\mddefault}{\updefault}{$Q'$}%
}}}}
\put(5671,-1689){\makebox(0,0)[lb]{\smash{{\SetFigFont{8}{9.6}{\rmdefault}{\mddefault}{\updefault}{$Q'$}%
}}}}
\put(5671,-55){\makebox(0,0)[lb]{\smash{{\SetFigFont{8}{9.6}{\rmdefault}{\mddefault}{\updefault}{$Q'$}%
}}}}
\put(4064,-59){\makebox(0,0)[lb]{\smash{{\SetFigFont{8}{9.6}{\rmdefault}{\mddefault}{\updefault}{$Q'$}%
}}}}
\put(6899,-1701){\makebox(0,0)[lb]{\smash{{\SetFigFont{8}{9.6}{\rmdefault}{\mddefault}{\updefault}{$Q'$}%
}}}}
\put(8506,-1701){\makebox(0,0)[lb]{\smash{{\SetFigFont{8}{9.6}{\rmdefault}{\mddefault}{\updefault}{$Q'$}%
}}}}
\put(8506,-67){\makebox(0,0)[lb]{\smash{{\SetFigFont{8}{9.6}{\rmdefault}{\mddefault}{\updefault}{$Q'$}%
}}}}
\put(6899,-71){\makebox(0,0)[lb]{\smash{{\SetFigFont{8}{9.6}{\rmdefault}{\mddefault}{\updefault}{$Q'$}%
}}}}
\put(2836,-55){\makebox(0,0)[lb]{\smash{{\SetFigFont{8}{9.6}{\rmdefault}{\mddefault}{\updefault}{$Q'$}%
}}}}
\put(2836,-1689){\makebox(0,0)[lb]{\smash{{\SetFigFont{8}{9.6}{\rmdefault}{\mddefault}{\updefault}{$Q'$}%
}}}}
\put(1229,-1689){\makebox(0,0)[lb]{\smash{{\SetFigFont{8}{9.6}{\rmdefault}{\mddefault}{\updefault}{$Q'$}%
}}}}
\put(1229,-59){\makebox(0,0)[lb]{\smash{{\SetFigFont{8}{9.6}{\rmdefault}{\mddefault}{\updefault}{$Q'$}%
}}}}
\end{picture}%

\caption{String-string scattering, leading to shortening of the strings and energy dissipation.}
}

We first estimate the rate of the relaxation as a result of the interactions with the thermal bath $gb$'s as follows:

At the $SU(N')$ phase transition, $T \sim T' \sim \Lambda'$, the energy density of the glueballs is roughly the same as in the unconfined phase, $T'^4 \sim \Lambda'^4$. The $Q'$ at the strings' ends interacts with the ambient glueballs with a cross-section
\begin{equation}
    \sigma_{int} \sim {\Lambda}'^{-2}
\end{equation}
(which is also the geometric cross-sectional area of the flux tube). All glueballs encountered in a time $\Delta t$ are then given a common translational ``drift'' speed $v_{Q'}$ and the total longitudinal momentum lost during this time interval is
\begin{equation}\label{eq:dp}
    \Delta p \sim [\Delta t \cdot v_{Q'} \cdot \Lambda'^{-2}] \Lambda'^4 \cdot v_{Q'},
\end{equation}
where the term in the brackets is the volume swept and the second term is the momentum given to a unit volume. To estimate $v_Q^2$ we use the virial theorem for linear potentials stating that the average kinetic energy of the non relativistic $Q'$s is half the average potential energy:
\begin{equation}
    T_{Q'}=\frac{1}{2}Mv_{Q'}^2 = \frac{1}{2}V \sim \Lambda'^2 \cdot L,
\end{equation}
with $L$ being the strings' length. From eq. (\ref{eq:dp}) we can write
\begin{equation}
    \frac{dp}{dt}=-\Lambda'^2 v_{Q'}^2.
\end{equation}
Multiplying by $p/M$, the left hand side becomes:
\begin{equation}
    \frac{p}{M} \frac{dp}{dt}=\frac{dE}{2dt},
\end{equation}
and since
\begin{equation}
    L(t)=\frac{E(t)}{\Lambda'^2},
\end{equation}
we obtain:
\begin{equation}
    \frac{dE}{dt}=-\Lambda'^2 \left(\frac{E}{M}\right)^{3/2}.
\end{equation}

Integrating over $dt$ between the initial $E_i \sim f\cdot {\Lambda}'$, and the final $E_f={\Lambda}'$ we obtain:

\begin{equation}\label{eq:relax_i}
    t^{relaxation}_i \sim 2 \frac{M^{3/2}}{\Lambda'^{5/2}} \sim (4 \cdot 10^{-10}-4) \ \mathrm{sec}
\end{equation}
for $\Lambda' \sim 10-10^5 eV$.

The relaxation time is very short in comparison with the Hubble time of $10^{10}~-~10^8$ sec required for the CMB temperatures to go through the relevant temperatures of $T~\sim~10$~eV and $T~\sim~100$~eV respectively.

The ambient glueballs also collide with the whole length of the extended flux tube. The transverse modes thereby excited, which de-excite via looping out and excised into $g'$-balls, do not, however, dissipate the longitudinal momentum/energy of the heavy $Q'$s which is of interest here.

We next estimate the relaxation time via the second, string-string collisions mechanism. With a reconnection probability of order $1$ \cite{Albrecht:1989mk,Hindmarsh:1994re,Polchinski:2006ee}. The time required for the first collision and rearrangement is:
\begin{equation}
    t_1 \sim (n_{string} \cdot \sigma_{string,string} \cdot v_{string})^{-1}.
\end{equation}

Using
\begin{equation}
    \sigma_{string,string} = \left(\frac{f}{\Lambda'}\right)^2
\end{equation}
 for the string string cross-section, a cross-section proportional to the square of the initial string length,
 \begin{equation}
    n_{string} \sim n_{Q'} \sim  \left(\frac{\Lambda'}{f}\right)^3
 \end{equation}
for the number density of strings and
\begin{equation}
    v_{string} \sim \left(\frac{T}{M}\right)^{1/2}
\end{equation}
for the relative speed of the centers of any two strings yields a rate:
\begin{equation}
    (t_1)^{-1}= \frac{\Lambda'^{3/2}}{f \cdot M^{1/2}}.
\end{equation}

If in such a collision the size of the strings is reduced, on average, by a factor $r$, we need to iterate this on average $k$ times such that $r^k=f$ for the final length of the string to be
\begin{equation}
    L_k \sim \Lambda'^{-1},
\end{equation}
which, by definition, is $f$ times smaller than the initial value $L_1$. The geometric string-string cross-section is smaller by $r^{-2}$ for the r-fold shorter strings. Hence the times between collisions in subsequent generations increase like $r^{2k}$. The total time required to relax via this mechanism to the final
\begin{equation}
    L_f=L_k \sim \Lambda'^{-1}=\frac{L_1}{f}
\end{equation}
is therefore
\begin{equation}
    t^{relaxation}_{ii}= {\Sigma}t_i \sim t_k \sim f^2 t_1 = 10^5 - 0.1 \ \mathrm{sec}
\end{equation}
for $\Lambda'= 10eV-10^5 eV$ and the above $f=2 \cdot 10^{5}$. While shorter than the relevant $10^{10} -100$ sec Hubble times it is much longer than $t^{relaxation}_{i}$ (eq. (\ref{eq:relax_i})), and hence is less important.

Once the flux tube's length becomes $L\ \sim \Lambda'^{-1}$ namely similar to its width, the system becomes spherical and the potential becomes, due to $g'$ exchange, Coulombic rather than linear. At this distance scale we have by definition,
\begin{equation}
    \alpha' \sim 1 \gg \alpha_{em} \sim \frac{1}{137},
\end{equation}
and the $Q' -\bar{Q'}$ attraction due to $g'$ exchange overcomes the small electric repulsion $\sim~2/9~\alpha_{em}$ between the $+2/3e$ charged $u$ and the $+1/3e$ charged $ud$ in the $\bar{Q'}u$ and $Q'ud$ respectively. Despite a logarithmic decrease with decreasing distance, the $g'$ attraction dominates at all scales. Note that the initial Bohr orbit with $r_n =L \sim \Lambda'^{-1}$ has a large $n$:
\begin{equation}
    n \sim \left(\frac{M}{\Lambda'}\right)^{1/2} \sim 3\cdot 10^3 - 3\cdot 10^5,
\end{equation}
and a classical description is appropriate.

Next, the $Q'ud$ baryons and $\bar{Q'}u$ mesons bound via $SU(N')$ forces rearrange into $Q'\bar{Q'}$ and a proton with large, hadronic $\sigma_H \sim GeV^{-2}$ cross-section. Let us estimate the average time required for that. Classically, the heavy meson and heavy baryon oscillate within their bound state. The probability of the above rearrangement occurring in each oscillation is:
\begin{equation}
    p_R \sim \frac{\sigma_H}{L^2},
\end{equation}
the ratio of the hadronic cross-section and the size of the bound state $L^2$.

Conservatively we take $L \sim \Lambda'^{-1}$ as the initial size. The oscillation frequency is conservatively estimated to be $v/L$ by neglecting downward cascading to smaller states with shorter periods, and using
\begin{equation}
    v \sim \left(\frac{T'}{M}\right)^{1/2} \sim \left(\frac{\Lambda'}{M}\right)^{1/2}.
\end{equation}

Thus the rate of rearrangements is
\begin{equation}
    p_R \cdot \frac {v}{L} \leq \frac{\sigma_H \cdot \Lambda'^{7/2}}{M^{1/2}}.
\end{equation}

This yields rearrangement times of $2 \cdot 10^5-2\cdot 10^{-9}$ sec for $\Lambda'= 10-10^5$ eV which again are much shorter than the corresponding Hubble times.

Once the $Q'\bar{Q'}$ QCD bound system forms, repeated emissions of $g'$, which as emphasized in section \ref{subsec:cosmological} above are perturbative on the relevant small (less than a Fermi) scale of the Quirkonium, relax it in a time of $ \sim 10^{-17}$ sec to the $S$-wave state, and annihilation into ordinary QCD gluons follows. Hadronization and decay of these generate photons. We next argue that such photons will have no observable effect even for the smallest $\Lambda' \sim 10$ eV contemplated with the annihilations occurring latest at
\begin{equation}
    t_{Hubble}=10^{10} \ \mathrm{sec},
\end{equation}
when the temperature is $T=10$ eV. The energy released in these annihilations constitutes only a small fraction,
\begin{equation}
    \frac{n_{Q'}}{s}\cdot \frac{M_{Q'}}{T} \sim  10^{-5},
\end{equation}
of the total radiation energy at this time. All emitted high energy photons quickly reach equilibrium by producing $e^+e^-$ pairs on the background. Photons with slightly lower energies, below the GKZ like threshold for $e^+e^-$ production, can still have multiple scattering on the background photons with $10^{14}$ cm$^{-3}$ densities with Delbruck cross-sections of
\begin{equation}
    \sigma_D \sim \frac{\alpha^4}{\mathrm{MeV}^2} \sim 10^{-31}\ \mathrm{cm}^2.
\end{equation}
Recall that the Hubble radius $R_H$ at $T \sim$ eV is roughly $ 10^{-4}$ times smaller than the present value, i.e $10^{24}$ cm namely approximately seven orders of magnitude larger than the mean free path for Delbruck scattering.

Finally, the $\sim 0.1-0.5$ TeV neutrinos from the annihilations red-shift at present to energies of $ \sim 100-500$ MeV and with fluxes $\sim 10^{-17}$ of the ordinary photon flux, namely to $10^{-4}$~cm$^{-2}$~sec$^{-1}$, way below the $\mathcal{O}(1)$ atmospheric neutrino fluxes. Hence, no striking signature of these ``very late'' $Q'-\bar{Q'}$ annihilations is expected.

\subsection{Baryonic States} \label{subsec:baryonic}
The discussion in section \ref{subsec:estimate} above suggests that all the $Q'$s annihilate leaving no relics. There is, however, a subtlety peculiar to the non-abelian $SU(N')$ confining gauge interactions that needs to be addressed. At the time of $SU(N')$ confinement, not only $Q'\bar{Q'}$ $SU(N')$ singlet mesons form but also baryon-like states made of $N'$ $Q'$s with all the $N'$ strings emanating from the $Q'$s joining at a junction (actually the ``$Q'$s'' refer at this point to $Q'ud$, but, as emphasized before this has no important effect).

This radically change our conclusions above if $N'=2$. The $N'=2$ representation of $SU(2)$, is self adjoint. Hence, the same $g'$ exchange forces act between particles and anti-particles, and the same string connects $Q'\bar{Q'}$ and $Q'Q'$ . Thus the $Q'Q'$ and $Q'\bar{Q'}$ states are degenerate and half of all $SU(2)$ confined systems formed initially are $Q'Q'$ or $\bar{Q'}\bar{Q'}$.

What is the eventual fate of the $Q'ud-Q'ud$ and $\bar{Q'}u-\bar{Q'}u$ $SU(N')$ confined states? The first step detailed above, where the initially long $SU(N')$ strings shrink to spherical bound states with radius $L\sim 1/\Lambda'$, is the same for $Q'Q'$ states and $Q'\bar{Q'}$ mesons. The $SU(3)$ color induced rearrangement reactions do \emph{not}, however, yield here a proton and Quirkonium as mentioned in section \ref{subsec:estimate} but rather
\begin{equation}
    Q'ud+Q'ud \rightarrow Q'Q'd + \mathrm{proton},
\end{equation}
and after $d\bar{d}$ and $u\bar{u}$ pair creation also
\begin{equation}
    \bar{Q'}u+\bar{Q'}u \rightarrow \bar{Q'}\bar{Q'}\bar{u} +\mathrm{proton} + \pi^+.
\end{equation}
The doubly heavy $Q'Q'q$ baryons and anti-baryons then quickly cascade via $g'$ emission to their ground states.

The $Q'Q'$ diquark ground state tightly binds by a Coulombic $SU(3)_c$ gluon to an $SU(3)_c$ triplet. For ordinary diquarks the two different flavors $u$ and $d$ fix the statistics in the color and spin anti-symmetric representations. Here the role of the two flavors is played by the two different colors of the new $SU(2')$ coupling to an $SU(2')$ singlet just like the $I=0$ light $ud$ diquark.

An additional light quark (which after a $d\rightarrow u$ $\beta$ decay becomes an up ($u$) quark) is required to make the heavy-heavy-light baryon a color singlet. Thus, the final $Q'Q'u$ state (and $\bar{Q'}\bar{Q'}\bar{u}$) are \emph{fractionally} charged hadrons. All stringent bounds on fractional charges and/or heavy isotopes apply excluding this possibility.

This then excludes $N'=2$ leaving us with $N'=3$, the maximum allowed by big bang nucleosynthesis considerations. As we show next this avoids all difficulties even if a non-negligible fraction, $f_B$ of the $Q'$s (or $\bar{Q'}$s) combine at the time of $SU(N')$ confining phase transition to form $Q'Q'Q'$ baryons or anti-baryons. (Again, ``$Q'$'' refers to $Q'ud$ and ``$\bar{Q'}$'' to $\bar{Q'}u$).

The first relaxation mechanism discussed in section \ref{subsec:estimate} above, via interaction of the strings with the $gb$'s, shrinks all three string bits in the baryon to $R\sim \mathcal{O}(1/\Lambda')$ in the short time intervals indicated above. The rearrangements analogous to the $SU(2')$ case above, into genuine $\bar{Q'}\bar{Q'}\bar{Q'}$ and a $uuu =\Delta^{++}$ or $Q'Q'Q'$ and a $uud$ and $ddu$ proton and neutron are somewhat slower, and one may wonder if prior to that we could actually have annihilation of the $Q'^3$ baryons and anti-baryons\footnote{Annihilation of the elementary $Q'$s is usually suppressed by $1/{M_Q'^2}$ factors. Here, however, the annihilation is actually a rearrangement into three $Q'\bar{Q'}$ mesons and just like baryon anti-baryon annihilations in QCD is likely to have large geometric cross-sections proportional to the size of the $Q'^3$ baryons. This size was $\sim \Lambda'^{-2}$ prior to rearrangement and formation of the QCD dominated quarkonium states.}\cite{Steigman:1976ev}.

\emph{If} these rearrangements were slower than the Hubble expansion rate $\sim\Lambda'^2/{M_{Planck}}$ (an issue to which we will return shortly), we will still have $Q'^3 - \bar{Q'}^3$ annihilations during the corresponding hubble times. With $f_B$ defined by:
\begin{equation}\label{eq:fraction}
    \frac{n_{Q'^3}}{s} \sim f_B\cdot 10^{-17},
\end{equation}
and (exothermic) annihilation cross-section
\begin{equation}
    \sigma_{ann}\sim\Lambda'^{-2}/\beta,
\end{equation}
the annihilation rate:
\begin{equation}
    n\cdot \sigma \cdot v \sim \frac{f_B \cdot 10^{-17} \cdot \Lambda'^3}{\Lambda'^2}
\end{equation}
equals the rate of Hubble expansion for $\Lambda'=10^5-10$ eV if
\begin{equation}
    f_B \sim 10^{-6} - 10^{-10},
\end{equation}
implying a small residual
\begin{equation}
    \frac{n_{Q'^3}}{s}\sim 10^{-23} - 10^{-27}.
\end{equation}

Let us next estimate the rearrangement rate of
\begin{equation}
    \left(Q'\bar{u}\right)^3 \rightarrow \bar{Q'}^3+\Delta^{++}.
\end{equation}

This rearrangement can proceed in two steps. First, a two body rearrangement of
\begin{equation}
    \bar{Q'}u +\bar{Q'}u \rightarrow \bar{Q'}^2 u^2,
\end{equation}
where the object on the right hand side is a color and color' Tetra-quark singlet,
followed by a quicker second rearrangement with the remaining $\bar{Q'}u$ into the final $\bar{Q'}^3$ and $u^3$ with the $u^3=\Delta^{++}$. We expect the $uu$ color anti-triplet to be less bound (by $(m_{\Delta}-m_N) \sim 300$ MeV) than the $ud$ diquark (where $m_N$ is the mass of a nucleon). We therefore need the heavy $\bar{Q'}\bar{Q'}$ system to be bound by more than that, and using a Coulombic binding,
\begin{equation}
    \frac{\alpha_s^2 \cdot m_{Q'}}{8n^2} > 0.3\ GeV,
\end{equation}
and $\alpha_s \sim 0.12$ appropriate to these scales, we need $n \sim \mathcal{O}(2)$ and hence, the $Q'$s have to get to within the corresponding Bohr radius
\begin{equation}
    \frac{1}{(m_{Q'}/2)(\alpha_s/2n)} \sim \frac{1}{15}\ GeV^{-1}
\end{equation}
a value which is $\sim 30$ times smaller than $0.2$ Fermi used in section \ref{subsec:cosmological} above.

The corresponding rearrangement times will therefor be $ 30^2\sim 10^3$ times longer, yet still shorter than the corresponding hubble times. Thus, we do not have very extensive annihilations of the larger $Q'^3$ baryons before the latter rearrange into the ``bare'' $Q'^3$'s.

The ``bare'' $Q'^3$ and $\bar{Q'^3}$ remaining after this stage are electrically neutral and, as we argue next, are unlikely to bind to nuclei. Hence, the very stringent upper bounds on fractionally charged and/or ultra heavy isotopes do not limit their relic density.

The ground state of the $Q'^3$ baryons has one $Q'Q'$ with $L=1$, unlike ordinary heavy baryons (for instance $bbb$) with purely $S$-wave ground states. This is due to the Fermi-Dirac statistics which motivated color in he first place: in order to be $SU(3)_c$ \emph{and} $SU(3')$ singlets, the three $Q'$s are antisymmetric in both color and color' and hence, should be antisymmetric in joint spin and orbit degrees of freedom. The three spin $1/2$ particles cannot be completely antisymmetrized. Rather, as in the nucleon, we have the $Q'Q'$ pairs half of the time in $S=0$ (and $L=0$) and otherwise in the symmetric $S=1$ and hence, in the antisymmetric $L=1$ state. This makes the $Q'^3$ ground state larger, enhancing interactions with nucleons.

These interactions, however, still fall short of generating $Q'$-nucleon bound states. The interaction between the color singlet $Q'^3$ and the nucleon is reminiscent of the Casimir-Polder interaction between neutral atoms \cite{Casmir:1947hx}. The latter is proportional to the product of the polarizeabilities of the two systems, which at most can be the corresponding volumes. The $m_N \sim 940$ MeV mass of the nucleon is almost entirely generated by the $\left(G_{\mu\nu}\right)^2$ gluonic condensate and the $\bar{\psi}\psi$ chiral condensate. When the heavy $Q'$ baryon is inside the nucleon it occupies a fraction
\begin{equation}
    f' \sim \left(\frac{r_{Q'^3}}{r_N}\right)^3
\end{equation}
of the nucleons' volume with $r_{Q'^3}$ the radius of the (small) $Q'^3$ QCD dominated baryon and $r_N$ the radius of the nucleon. The above condensates are modified in the presence of the chromoelectric fields inside the $Q'^3$ baryon, reducing the nucleons mass and causing attraction (like the Casimir Polder interaction in QED). An extreme assumption, maximizing this interaction, is that the contribution to the nucleon mass from the above region vanishes. This then yields a potential
\begin{equation}
    U=- f'\cdot m_N
\end{equation}
and a range $r_N$. Since at least two gluons need to be exchanged, we have the additional product
\begin{equation}
    \alpha_{s}\vert_{Nucleon}\cdot\alpha_{s}\vert_{Q'^3}
\end{equation}
of the strong coupling at the nucleon and quarkonium scales, amounting to another factor of
$\sim 1/30$.

The ground (and $L=1$) states of $Q'^3$ are in the Coulombic regimes of $SU(3)_c$. Approximating the system as a $Q'^2$ diquark in an $L=1$ state with another $Q'$, and using the nominal $m_{Q'}=1$ TeV, we find that $r_{n=2}$ corresponding to the $L=1$ assumed, is:
\begin{equation}
    r_{n=2}=\frac{n^2}{\mu\alpha_{s}/2} \sim 2\cdot 10^{-15}\ \mathrm{cm},
\end{equation}
where we have used a reduced mass
\begin{equation}
    \mu=\frac{2}{3}m_{Q'}
\end{equation}
and a QCD coupling $\alpha_{s}\sim 0.12$ appropriate to the relevant distance scale. The extra $1/2$ appears in the denominator due to the corresponding reduction of the interaction between two $3_c$ quarks to yield a $\bar{3}_c$ as compared with the interaction between $Q'$ and $\bar{Q'}$ in color singlet quarkonium.

To this we need to add (in quadrature) the radius of the initial $S$-wave diquark, which is about half as large to have
\begin{equation}
    r_{Q'^3}\sim 2.5 \cdot 10^{-15}\ \mathrm{cm}.
\end{equation}
Using $r_N \sim 1.4 \cdot 10^{-13}$ Fermi this yields\footnote{We utilize this interactions for large nuclei with radii $R(A,Z)\sim 1.4 \cdot A^{1/3}$ Fermi. The large penalty of kinetic (uncertainty) energy prevents the $Q'^3$'s from concentrating inside individual nucleons. Hence we have the $Q'^3$ bound to the whole large nucleus and $1.4$ Fermi in the expression for the nuclear size appears.}
\begin{equation}
    U \sim -5\ \mathrm{KeV}\cdot 1/{30}=0.15\ \mathrm{KeV}.
\end{equation}

The condition that a heavy $m(A,Z) \sim A \cdot m_N$ nucleus binds to $Q'^3$ is:
\begin{equation}
    \left(2m(A,Z) \cdot |U|\right)^{1/2}\cdot R(A,Z)> \pi/2
\end{equation}
which becomes
\begin{equation}
    A>340,
\end{equation}
and we have no nuclei big enough to bind.

The above $Q'^3-(A,Z)$ interaction causes relic $Q'^3$'s moving with virial velocities of the order of $10^{-3}$ to scatter elastically on a Germanium ($A \sim 75$) nuclei, for instance, in underground detectors. The Born approximation, applicable for such weak interactions yields
\begin{equation}
    \sigma_{elastic} \sim 10^{-34}\ \mathrm{cm}^2.
\end{equation}
The present best bound from CDMS2 (expressed in terms of cross-sections for WIMP-single nucleon scattering) for TeV WIMP relics constituting the galactic halo is approximately $10^{-42}$ cm$^2$. This translates\footnote{There is an $A^2$ coherence factor and another $A^2$ factor arises from the reduced mass which is approximately that of the nucleus.} to a cross-section which is $75^4\sim 3 \cdot 10^7$ times weaker on Germanium, thus allowing the $Q'^3$ to constitute $\sim 1/3$ of the halo mass. Since we know already that the early $Q'^3$ annihilations alone reduce their densities far bellow that, no new constraint on $f_{B}$, the fraction of $Q'$s surviving inside $Q'^3$, arises.

In High energy collisions and most likely in QCD phase transition, baryon-anti-baryon production is suppressed by $\mathcal{O}(1/10-1/100)$ relative to mesons. Since in the present case the produced $Q'^3$ baryons (and anti-baryons) are to start with $\sim 10^6$ times larger than the $SU(N')$ scale, there is a high probability that the junction points, where the three strings join to form such a baryon and in an anti-baryon, which, as emphasized in section \ref{subsec:estimate} above, constitute light and fast degrees of freedom, will overlap and annihilate yielding a final $Q'^3$ density well below the detectability threshold.

Before concluding this section, we note that the cross-sections on Germanium (and interaction potentials) of the $Q'^3$'s are much larger than for technibaryons containing colored constituents calculated some time ago \cite{Chivukula:1992pn} to be $\sim 10^{-41}$ cm$^2$. This reflects the much smaller $\mathcal{O}\left( \ \mathrm{TeV}^{-1}\right)$ radius of the technibaryons as compared with those of $Q'^3$.

\subsection{Quirks at LHC}\label{subsec:LHC}
Our main conclusion is that extending the standard model with a new $SU(3')$ and with new massive, stable $Q'$s transforming as $(3',3_c)$ is cosmologically viable and consistent with all bounds, if $SU(3')$ confines at a scale $\Lambda'$ in the $10$ eV to $10^5$ eV range. While this model may ruin the nice feature of gauge coupling unification, this consistency is non trivial. Indeed, generic standard model extensions, which keep coupling unification tend to have unconfined, fractionally charged, particles \cite{Raby:2007hm}. If not very heavy, these constitute dangerous relics which can be ruled out by cosmological considerations and the experimental bounds \cite{Steigman:unpublished}. Not everything that is possible necessarily happens, yet we are encouraged to consider possible signatures of the putative new particles and interactions at the LHC collider\footnote{Such considerations have been made by M. Luty.}.

To estimate the production cross-section $\sigma_{Q'}$ of $Q'\bar{Q'}$ with $M_{Q'}=1$ TeV at LHC (at center of mass energy $W=14$ TeV), we use the measured $\sim 8$ picobarns production cross-section of top-anti-top pairs (with $m_t=180$ GeV, $\sim 5.5$ times lower than that of a TeV $Q'$) at the Fermi-lab collider (with $7$ times lower $\sim 2$ TeV center of mass energy). If naive scaling could be applied, we would expect, accounting for the color' index of $Q'$s, that
\begin{equation}
    \sigma_{Q'}|_{LHC} \sim \cdot 3 \cdot \sigma_t|_{\mathrm{Tevatron}} \sim 20\ \mathrm{picobarns}.
\end{equation}

With $\sigma_ {total}(pp)|_{LHC} \sim 100$ mb we expect $Q'\bar{Q'}$ pair production in about $\sim 3$ out of $10^{10}$ collisions in LHC.

QCD corrections cause substantial deviation from scaling. These are partially accounted for by the increasing cross-section and also by the fact that the energy to mass ratio is $7/{5.5}\sim 1.3$ times higher in LHC, making the above estimate plausible.

For the maximal LHC luminosity of $L\sim 10^{34}$ cm$^{-2}$sec$^{-1}$ we expect $0.3$ $Q'$ production events among the $10^9$ proton-proton collisions occurring each second. Can one pick up this tiny $Q'\bar{Q'}$ signal?

There is an appreciable probability of $\sim 2\alpha_{s} \sim 0.2$ that the hard scattering producing $Q'\bar{Q'}$ at $s^2\gtrsim 4$ TeV$^2$ is accompanied by another gluon jet with substantial transverse energy deposition
\begin{equation}
    E>\mathcal{O}(50)\ \mathrm{GeV}
\end{equation}
in the hadronic calorimeters. This trigger may thus yield a sample enriched in $Q'$ production events.

The $Q'\bar{q}$ and $\bar{Q'}q$ mesons have large, $\sigma_{Q'\bar{q}-N}\sim 10$ mb, cross-sections on nucleons. Yet, due to the very large (more than $1000$) ratio of $M_Q'$ and the $\lesssim$ GeV scale of the light quarks, these mesons have tiny (less than $10^{-3}$) inelasticities. Hence, the mesons can suffer more than $1000$ hadronic collisions and the mean free path of such a meson is
\begin{equation}
    l_{mfp}=\frac{1}{n_p\cdot \sigma_{Q'\bar{q}-N}}
\end{equation}
and thus the mesons traverse $\sim 10^5$ gr/cm$^2$ before stopping. The $2/3$ and $1/3$ charged $\bar{Q'}u$ and $Q'ud$ also loose energy by ionization. For a mildly relativistic $Q'$, with $\beta_{Q'} \lesssim 0.7$, these losses are less than MeV$/($gr$\cdot$cm$^2)$ in iron. Thus, a $Q'$ with initial kinetic energy of $K \sim 1/4$ TeV reaches and traverses all muon detectors. If $K^*$ is the kinetic energy of the relative motion in the $Q'\bar{Q'}$ center of mass frame, the string between $Q'$ and $\bar{Q'}$ can stretch (in this frame) to a length of
\begin{equation}
    L' = \frac{K^*}{\sigma} \sim \frac{K^*}{\Lambda'^2},
\end{equation}
with $\sigma$ being the string's tension. Table \ref{fig:length} shows the length of the string for $K^*=1/2$ TeV and possible values of $\Lambda'$.

\TABULAR{|p{2cm} |p{2cm}|}{
    \hline
    $\Lambda'[\mathrm{eV}]$ & $L'[\mathrm{meters}]$\\
    \hline
    $10$ & $10^3$ \\
    $100$ & $10$ \\
    $10^3$ & $0.1$ \\
    $10^4$ & $10^{-3}$\\
    $10^5$ & $10^{-5}$\label{fig:length}\\
        \hline}
        {The string's length for possible values of $\Lambda'$ and $K^*~=~1/2$~TeV.}

After reaching the maximal extension, the string will bring the $Q'$ and $\bar{Q'}$ close together again. In vacuum, the system could oscillate almost indefinitely as the probability of annihilation in each crossing $\sim (\Lambda'/M_{Q'})^2$ is less than $10^{-20}$.
The Schwinger mechanism for breaking the string via $Q'\bar{Q'}$ pair production is suppressed by $e^{-(M/\Lambda')^2}$. One may still wonder if $g'g'$ glueball emission or excision of small closed loops at any point of the long $SU'(3)$ string cannot dissipate the latter in a short time of order $1/{\Lambda'}$. This is not the case as the following argument shows: Unlike the long floppy strings forming in a thermal environment around the time of $SU'(3)$ phase transition, the string connecting the LHC produced $Q'$ pair is straight and taut during all phases of the oscillation described above.

Indeed, the extended string represents at all times the true ground state of the light $SU(3')$ degrees of freedom given the location of the heavy Quirks. Furthermore, apart from the regions very near to the receding (or converging) end Quirks, the whole length of the string/flux tube is stationary. Therefore, only the end region could, in principle, radiate.

A conservative overestimate of the rate of radiation is provided by treating the end Quirks as unconfined carriers of the $SU(3')$ charge. Using the Larmor formula for $dW/dt$ (which is adequate for the slightly relativistic Quirks) we find that $\sim 100$ sec are required for dissipating the energy via this mechanism. This is far longer (by $10^7$) than the estimated stopping time (see eq. \ref{eq:stopping_time} below).

M. Luty originally suggested that particles at the far muon detectors, which rather than diverging away from the intersection point, converge towards each other will be a striking, unique, signature of the new Quirks and a low $\Lambda'\sim 100$ eV for which $L'~\sim~10$~meters. Unfortunately, trajectories which do not extrapolate back to the intersection point are likely to be cosmic rays background and are discarded in the present LHC experimental protocol.

Amusingly, for $\Lambda'\sim$ KeV, the two trajectories interwind sufficiently closely, so that the net braided trajectory may extrapolate to the intersection point, yet the $10$ cm separation could suffice for resolving the two individual trajectories in segmented detectors, leading again to a striking signal.

During the $\sim 10$ years optimal running of the LHC, the above estimate suggests that about $10^8$ $Q'\bar{Q'}$ pairs will be produced, a number comparable to or larger than that of the much more motivated and studied supersymmetric partners. It seems likely that should any new physics of the type discussed here be realized in nature, LHC may indicate its existence\footnote{Cosmic ray protons of energies $E>10^{17}$ eV (corresponding to center of mass energy of $14$ TeV as in LHC) also generate $Q'$s. Unfortunately, the flux of such protons is only $ \sim 2\cdot 10^{-10}$(m$^{-2}$ sec$^{-1}$)\cite{Yao:2006px}, yielding $2 \cdot 10^{-20}$ $Q'$s per (m$^2$ sec) and an area of a km$^2$ accumulates only $10^3$ $Q'$s in a billion years.}.

\subsection{The Fate of the Produced Quirks}\label{subsec:fate}
We argued at length that all early universe $Q'$ and $\bar{Q'}$s annihilate. Here, we would like to address the fate of the $Q'$s produced in LHC and/or in cosmic ray collisions at present. As indicated above, $Q'$ hadrons loose energy and eventually stop after traversing $\mathcal{O}(km)$ of earth. The $SU(3')$ string keeps pulling the $Q'$ and $\bar{Q'}$ towards each other and one might expect that all the $\bar{Q}'-Q'$ pairs will annihilate. As we show next this need not be the case.

Consider first $\bar{Q'}q$ heavy mesons. The stopping time while traversing a kilometer of earth,
\begin{equation}\label{eq:stopping_time}
    t_s \sim \frac{10^5\ \mathrm{cm}}{c} \sim 10^{-5}\ \mathrm{sec},
\end{equation}
is much shorter than the $\mathcal{O}(1)$ sec duration of $\bar{Q'}d$ beta decay into $\bar{Q'}u$. These $-1/3 e$ charged mesons are attracted by the Coulombic potentials
\begin{equation}
    U_{Coulomb}\sim \frac{(Z/3) \cdot \alpha_{em}}{R(A,Z)} \sim 1.2 \ \mathrm{and} \ 2\ \mathrm{MeV},
\end{equation}
existing at the nuclear radius for $A=20$, $Z=10$ silicon and $A=40$, $Z=20$ calcium nuclei. With negligible kinetic energies of the heavy bound nuclei
\begin{equation}
    E_k\sim \frac{h^2}{2\cdot M(A,Z)\cdot R(A,Z)^2},
\end{equation}
this presents a lower bound for the actual binding.

We would like to argue next that for an $S$-wave $\bar{Q'}d$ (or $\bar{Q'}u$) state localized around the nucleus in a thin $1$ Fermi shell at $R(A,Z)$, the nuclear radius, there is also a significant nuclear binding. As for $K^+$, $K^0$ and nucleons, we can model the $\bar{Q}'q$-nucleon interactions by exchanges of the non-strange $\sigma$ and $\omega$ mesons \cite{Serot:1984ey} (one pion exchange is parity forbidden since the ground state mesons are pseudoscalars). The high (kinetic) uncertainty penalty for localized Kaons smears a putative bound $S$-wave state over the whole nucleus. The expectation value of the potential energy is then:
\begin{equation}
    \left<V\right> \sim \int d^3r \left(V_{\omega}(r) +V_{\sigma}(r)\right).
\end{equation}

The repulsive $\omega$ exchange and the attractive $\sigma$ exchange largely cancel, leaving a small net repulsive potential energy, excluding binding, and yielding the observed repulsive scattering lengths. The attractive $\sigma$ exchange has, however, a longer $\sim 0.4-0.5$ Fermi range as compared with the $0.25$ Fermi for $\omega$ exchange. Hence a heavy $Q'\bar q$ meson, placed at a distance of $\mathcal{O}(\mathrm{Fermi})$ from the nuclear surface will experience a suppressed attractive potential, which is still $\sim 8$ times enhanced relative to the repulsive part. The net attraction of $\mathcal{O}(10)$ MeV will then directly contribute to the binding.

A beta decay of $d$ into the $+2/3$ charged $u$ quarks faces now a $2.5$ and $4$ MeV Coulomb barrier for $Z=10$ and $20$, and with $m_d-m_u-m_e \sim3$ MeV, a $Z>12$ nuclei is forbidden. Even if such decays happen in the lighter nuclei (with rather long $\sim 100$ sec lifetimes) the daughter nucleus is likely to still remain bound by nuclear interactions. We note that $\bar{Q'}u$ mesons can also hadronically bind to nuclei despite the Coulomb barrier in collisions with center of mass energy exceeding this barrier.

We next turn to $Q'$s. As noted in section \ref{sec:reduced} above, $Q'\bar{q}$-nucleon collisions rearrange the light quarks as in eq. (\ref{eq:baryon_formation}) and all $Q'$s form $Q'ud$ baryons. The latter have $+1/3$ charges and are Coulomb repelled by nuclei. The $Q'ud$ baryons \emph{can}, however, bind to nuclei before slowing down to $\mathcal{O}(2)-\mathcal{O}(3)$ MeV required to overcome the Coulomb barrier. The $Q'ud$ baryons have large hadronic $\mathcal{O}(30)-\mathcal{O}(40)$ MeV binding to $A \sim 20-40$ nuclei. This binding is inferred by realizing that $Q'ud$ is analogous to the $\Lambda(1115)$ Hyperon, $sud$ \cite{Dalitz:1958di,Povh:1978mx}, which has a deep binding potential in nuclei,
\begin{equation}
    V\sim -30\ \mathrm{to} -40\ \mathrm{MeV},
\end{equation}
for $(A,Z)$ ranging from $(20,10)$ to $(40,20)$ \cite{Vidana:1998ri}. Further kinetic energy effects are negligible for the $\sim 17-34$ times heavier nuclei. Thus, the system of $Q'ud$ and nuclei can sit in many angular momentum states and still be bound by this potential. Up to few MeV Coulombic correction, we expect binding energies in this range.

In order to form the bound state, the extra kinetic energy of relative motion has to be emitted via a photon or a pion:
\begin{equation}
    Q'ud +(Z',A') \rightarrow Q'ud (Z',A') + \gamma \ (\mathrm{or} + \pi).
\end{equation}

The first process is suppressed by $\alpha_{em}$ and both processes are suppressed by the ratio $B.E/ T_{Q'ud}$ where $T_{Q'ud}$ is the kinetic energy of the $Q'ud$-nucleus system in its center of mass. For the hadronic pion emission process, the kinetic energy had to exceed $m_\pi~\sim~140$~MeV. Still we have several $Q'ud$ nuclear collisions, where this binding can occur, so that assuming that approximately $20\%$ of all $Q'ud$ bind to nuclei is reasonable.

The bound $Q'ud(A,Z)$ or $\bar{Q'}d(A,Z)$ complexes loose their kinetic energies extremely fast due to heavy ionization and frequent atomic collisions, and stop, forming interstitial impurity in some grain. No annihilations of the $Q'$ and $\bar{Q'}$s can now happen as it is impeded by the very strong $\sim Z\cdot Z'$ Coulomb repulsion between the nuclei.

The positively ($2/3$ and $1/3$) charged $\bar{Q'}u$ and $Q'ud$ that have failed to bind to nuclei when their energies exceeded the Coulomb barrier, still bind to the atoms. These will be extended few Angstrom analogous to ordinary molecular bound states, yet with smaller bindings of $\sim 1/4-1$ eV because of the smaller charges (since the binding for closed shells is only via polarization of the wave-function of the $Z$ electrons it should scale with ${e_Q'}^2$.

The intersection points of the LHC collider are about $100$ meters underground and the range of $\sim$1/2 TeV $Q'$ containing heavy hadrons in ground material is approximately 1/2 kilometer\footnote{We assume that the $Q'$ is generally produced with a kinetic energy $T'$ which is a finite fraction ($0.1-0.5$) of its mass.}. For $\Lambda'>30$ eV the separation allowed by the confining strings for kinetic energy of $\sim 1/2$ TeV is
\begin{equation}
    L_{max} < 50 m \ \ \ \ \mathrm{(for}\ \Lambda'>30\ \mathrm{eV)}.
\end{equation}
We then expect that half of the $Q'-\bar{Q'}$ pairs produced will jointly move downwards where the individual $Q'$ and $\bar{Q'}$ particles will be captured (about $20\%$ into nuclei and the rest bound to atoms) and half will travel upward into space.

For $\Lambda'<30 $ eV we are likely to have one member of the produced pair move upwards but then be pulled back to the other $Q'$ which will be captured underground and eventually be captured as well.

Is it possible that the $Q'$ and $\bar{Q'}$ paired by a string connection will actually be fixed in the locations where the individual $Q'$ and $\bar{Q'}$ were trapped and be unable to move towards each other? For $\Lambda'\sim 10-100$ eV, the constant string tension force pulling the $Q'$s together is $0.05$ eV$/\AA$ and $5$ eV$/\AA$ respectively. If the tension is less than $\sim 0.5$ eV$/\AA$ this force may be too weak to overcome the chemical forces binding some of positively charged $\bar{Q'}u$ and $Q'ud$ to atoms, and will not be able to move these atoms to the nuclei of which the remaining $Q'$ and $\bar{Q'}$ hadrons are attached within the grains.

It is worth emphasizing that the $Q'$ numbers and in particular their concentrations vastly increase as $m_{Q'}$ decreases. For instance, if the latter mass is only $1/2$ TeV rather than $1$ TeV, its production cross-section should be $\sim 10$ times larger. The kinetic energy of the lighter $Q'$s will be on average $1/2$ to $1/4$ times smaller. This will make the range in earth correspondingly shorter and the final $Q'$ concentration will be between $\sim 10\cdot 2^3=80$ and $\sim 10\cdot 4^3=600$ times larger.

\subsection{The Physics of Trapped Quirks}\label{subsec:trapped}
In this final section we briefly speculate on the fascinating possibilities arising if the above scenario is realized and a clever segregation technique enables finding a grain containing one $Q'$ hadron. Such a grain experiences, in addition to known gravitational and electromagnetic forces, a ``mysterious'' additional constant force moving it towards the $Q'$ (or $\bar{Q'}$) at the other end of the string. The force,
\begin{equation}
    F=\sigma= \Lambda'^2= 1.6 \cdot 10^{-4} \ \mathrm{dyne},
\end{equation}
equals, for $\Lambda'\sim 30$ eV, to the gravitational force for a grain weighing $\sim 2\cdot 10^{-7}$ gr, and by careful experiments could be measured for much larger grains.

The above $\Lambda'$ is the maximal which still does \emph{not} tear the $Q'$s out of their host grains, so that we can take the grain and move it, leaving behind an $SU(3')$ long string.

To most dramatically illustrate this, consider having the grain in the $\sim 150$ kg Pioneer, presently at the edge of the solar system at about $100$ A.U $\sim 1.5 \cdot 10^{15}$ cm away. The gravitational acceleration due to the sun's gravitational field ($\sim 6\cdot 10^{-3}$ cm/s$^2$) is monitored and an anomalous enhancement of $1$ in $10^4$ has been reported. One string stretching between Pioneer and earth could account for it if $\Lambda > 10^3$ eV, a mere factor 30 beyond the maximally allowed value\footnote{If the Pioneer continuously tumbles around, as the WMAP satellite does, than the $Q'$ can be dynamically trapped there for much longer times as the direction of the required escape crack would be constantly changing...}.

For such an extension, the potential energy stored even in the nominal $0.5$ eV $/\AA$ string is $10^{23} eV \sim 10^{14}$ GeV. If we vaporize the grain holding our $Q'$ in space, the attraction towards the other partner in earth would cause acceleration to such energies! Thus, such strings could be the ultimate, perhaps even Trans-Planckian accelerators\footnote{The authors of Ref. \cite{Casher:1997rr} asserted that laws of microphysics and cosmology do not allow the attainment of such energies, however if the Quirk scenario is realized, it appears not to be the case.}.

 Since the $SU(3')$ string has no interaction with matter we could envision another striking situation were the two grains at the string's end are moved to antipodal points on the earth and yet when pulling on one end the other would respond within a time of order $R_E/c \sim 0.02$ sec!

\section{Conclusions}
In this paper we have shown that long lived massive colored particles, $\chi$, are consistent with cosmological bounds. The relic abundance of such particles is reduced below the detectability threshold in several stages of annihilation. First, annihilations via perturbative QCD occur. After the QCD phase transition, the massive colored particles form heavy mesons (or glueballinos) which have a large geometric scattering cross-section on each other yielding heavy $\chi\chi$ states with large angular momentum. These states then cascade down to lower angular momentum states and finally annihilate. Even though collisions with ambient pions can break the newly formed heavy-heavy states (and thus impede the annihilations), these collisions also cause further cascading to lower angular momentum states and finally to annihilation.

M. Luty suggested a Quirk model where, in addition to $SU(3)_c$, the heavy colored particles carry an additional $SU(N')$ non-abelian gauge group.
For $SU(N')$ scales much smaller than the mass of the particles, unbreakable flux tubes (or strings) of macroscopic size connect two such particles . The existence of such a gauge group has little effect on the first stage of annihilation. We have shown that almost all of heavy hadrons containing these heavy colored particles annihilate (so that the number of fractionally charged particles and anomalous heavy isotopes drops below detectability threshold) for $N'\geq~3$. Furthermore, since $N'>3$ is forbidden by big bang nucleosynthesis, the only gauge group consistent with all cosmological observations is $SU(3')$.

If the mass of the Quirks is indeed of the order of $1$ TeV, pairs of $Q'-\bar{Q'}$ will be created in LHC. The string between the two particles extends until it reaches its maximum length, pulls the Quirk and anti-Quirk close together again causing the particles to oscillate. Such events, which do not diverge away from the interaction point, will be discarded in LHC experiments unless the maximum length of the string is small enough so that the particles interwind sufficiently closely and still point to the intersection.

Quirks (either created in accelerators or in cosmic rays)
bind to nuclei in the earth and the electrostatic repulsion between the nuclei prevents their annihilations. For small enough values of $\Lambda'$ the attractive force between the Quirks will not be able to overcome the chemical forces binding the heavy baryons and mesons to atoms and nuclei, and the Quirk and anti-Quirk, bound by a string will be trapped in fixed locations and will not be able to move towards each other.

We have concluded by commenting, that if it were possible to isolate two grains connected by a long $SU(3)$ string, it would be feasible to measure the force the new gauge group exerts. Furthermore, separated to large enough distances and then released, the particles can accelerate to extremely high energies.

\acknowledgments
The above work has been conducted over the last year at Tel Aviv University and various other places. The authors would like to thank A. Casher, R. Furnstahl, S. Itzhaki, S. Raby for helpful comments and to G. Steigman for reading the manuscript and for many comments.

This work was largely inspired by M. Luty's Quirk model and his speculation on Quirk detectability at LHC.

\bibliographystyle{JHEP}

\bibliography{hadronic}

\providecommand{\href}[2]{#2}\begingroup\raggedright\begin{thebibliography}{10}

\bibitem{Arkani-Hamed:2004fb}
N.~Arkani-Hamed and S.~Dimopoulos, {\it Supersymmetric unification without low
  energy supersymmetry and signatures for fine-tuning at the {LHC}},  {\em
  JHEP} {\bf 06} (2005) 073,
  [\href{http://xxx.lanl.gov/abs/hep-th/0405159}{{\tt hep-th/0405159}}].

\bibitem{Kang:2006yd}
J.~Kang, M.~A. Luty, and S.~Nasri, {\it The relic abundance of long-lived heavy
  colored particles},  \href{http://xxx.lanl.gov/abs/hep-ph/0611322}{{\tt
  hep-ph/0611322}}.

\bibitem{Griest:1989wd}
K.~Griest and M.~Kamionkowski, {\it Unitarity limits on the mass and radius of
  dark matter particles},  {\em Phys. Rev. Lett.} {\bf 64} (1990) 615.

\bibitem{Wolfram:1978gp}
S.~Wolfram, {\it Abundances of stable particles produced in the early
  universe},  {\em Phys. Lett.} {\bf B82} (1979) 65.

\bibitem{Nussinov:1975mw}
S.~Nussinov, {\it Colored quark version of some hadronic puzzles},  {\em Phys.
  Rev. Lett.} {\bf 34} (1975) 1286--1289.

\bibitem{Gupta:1981ve}
S.~Gupta and H.~R. Quinn, {\it Heavy quarks and perturbative {QCD}
  calculations},  {\em Phys. Rev.} {\bf D25} (1982) 838.

\bibitem{Chen:1993tr}
H.~Chen, J.~Sexton, A.~Vaccarino, and D.~Weingarten, {\it Glueball mass
  predictions of the valence approximation to lattice {QCD}},
  \href{http://xxx.lanl.gov/abs/hep-lat/9308010}{{\tt hep-lat/9308010}}.

\bibitem{Vaccarino:1999ku}
A.~Vaccarino and D.~Weingarten, {\it Glueball mass predictions of the valence
  approximation to lattice {QCD}},  {\em Phys. Rev.} {\bf D60} (1999) 114501,
  [\href{http://xxx.lanl.gov/abs/hep-lat/9910007}{{\tt hep-lat/9910007}}].

\bibitem{Carlson:1992fn}
E.~D. Carlson, M.~E. Machacek, and L.~J. Hall, {\it Selfinteracting dark
  matter}, . HUTP-91-A066.

\bibitem{Raby:2007hm}
S.~Raby and A.~Wingerter, {\it Gauge coupling unification and light exotica in
  string theory},  {\em Phys. Rev. Lett.} {\bf 99} (2007) 051802,
  [\href{http://xxx.lanl.gov/abs/0705.0294}{{\tt 0705.0294}}].

\bibitem{'tHooft:1980xb}
{'t Hooft, Gerard, (ed.)} {\em et.~al.}, {\it Recent developments in gauge
  theories. proceedings, {Nato} advanced study institute, {Cargese}, {France},
  {August} 26 - {September} 8, 1979}, . New York, Usa: Plenum (1980) 438 P.
  (Nato Advanced Study Institutes Series: Series B, Physics, 59).

\bibitem{Albrecht:1989mk}
A.~Albrecht and N.~Turok, {\it Evolution of cosmic string networks},  {\em
  Phys. Rev.} {\bf D40} (1989) 973--1001.

\bibitem{Hindmarsh:1994re}
M.~B. Hindmarsh and T.~W.~B. Kibble, {\it Cosmic strings},  {\em Rept. Prog.
  Phys.} {\bf 58} (1995) 477--562,
  [\href{http://xxx.lanl.gov/abs/hep-ph/9411342}{{\tt hep-ph/9411342}}].

\bibitem{Polchinski:2006ee}
J.~Polchinski and J.~V. Rocha, {\it Analytic study of small scale structure on
  cosmic strings},  {\em Phys. Rev.} {\bf D74} (2006) 083504,
  [\href{http://xxx.lanl.gov/abs/hep-ph/0606205}{{\tt hep-ph/0606205}}].

\bibitem{Steigman:1976ev}
G.~Steigman, {\it Observational tests of antimatter cosmologies},  {\em Ann.
  Rev. Astron. Astrophys.} {\bf 14} (1976) 339--372.

\bibitem{Casmir:1947hx}
H.~B.~G. Casmir and D.~Polder, {\it The influence of reatardation on the
  {London}-van der {Waals} forces},  {\em Phys. Rev.} {\bf 73} (1948) 360--372.

\bibitem{Chivukula:1992pn}
R.~S. Chivukula, A.~G. Cohen, M.~E. Luke, and M.~J. Savage, {\it A comment on
  the strong interactions of color - neutral technibaryons},  {\em Phys. Lett.}
  {\bf B298} (1993) 380--382,
  [\href{http://xxx.lanl.gov/abs/hep-ph/9210274}{{\tt hep-ph/9210274}}].

\bibitem{Steigman:unpublished}
G.~Steigman, {\it On the severe bounds of fractionally charged uncolored stable
  remnants},  {\em Unpublished}.

\bibitem{Yao:2006px}
{\bf Particle Data Group} Collaboration, W.~M. Yao {\em et.~al.}, {\it Review
  of particle physics},  {\em J. Phys.} {\bf G33} (2006) 1--1232.

\bibitem{Serot:1984ey}
B.~D. Serot and J.~D. Walecka, {\it The relativistic nuclear many body
  problem},  {\em Adv. Nucl. Phys.} {\bf 16} (1986) 1--327.

\bibitem{Dalitz:1958di}
R.~H. Dalitz and B.~W. Downs, {\it Hypernuclear binding energies and the
  {Lambda}-nucleon interaction},  {\em Phys. Rev.} {\bf 111} (1958) 967--986.

\bibitem{Povh:1978mx}
B.~Povh, {\it Hypernuclei},  {\em Ann. Rev. Nucl. Part. Sci.} {\bf 28} (1978)
  1--32.

\bibitem{Vidana:1998ri}
I.~Vidana, A.~Polls, A.~Ramos, and M.~Hjorth-Jensen, {\it Binding energy of
  {Lambda} hypernuclei from realistic {Y N} interactions},
  \href{http://xxx.lanl.gov/abs/nucl-th/9809060}{{\tt nucl-th/9809060}}.

\bibitem{Casher:1997rr}
A.~Casher and S.~Nussinov, {\it Is the {Planck} momentum attainable?},
  \href{http://xxx.lanl.gov/abs/hep-th/9709127}{{\tt hep-th/9709127}}.

\end{thebibliography}\endgroup

\end{document}